\documentstyle[11pt]{article}
\let\LARGE=\Large
\let\Large=\large
\let\large=\normalsize
\newcommand{\eqn}[1]{(\ref{#1})}

\renewcommand{\d}{\delta}
\newcommand{\pa}{\partial}

\newcommand{\e}{\epsilon}

\newcommand{\m}{\mu}

\newcommand{\n}{\nu}

\newcommand{\ft}[2]{{\textstyle\frac{#1}{#2}}}
\newcommand{\be}{\begin{equation}}
\newcommand{\eq}{\end{equation}}

\def\beqa{\begin{eqnarray}}
\def\eeqa{\end{eqnarray}}

\begin {document}
\begin{titlepage}
\begin{center}
\hfill THU-99/09\\
\hfill {\tt hep-th/9904005}\\

\vskip 3cm

{ \LARGE \bf Deviations from the Area Law for Supersymmetric Black Holes}

\vskip .3in

{\bf Gabriel Lopes Cardoso$^{1}$\footnote{\mbox{
\tt 
cardoso@phys.uu.nl}}, 
Bernard de Wit$^{2}$\footnote{\mbox{
\tt 
bdewit@phys.uu.nl}} 
and Thomas Mohaupt$^3$\footnote{\mbox{
\tt 
mohaupt@hera1.physik.uni-halle.de}}}
\\

\vskip 1cm


{\em

\centerline{$^{1,2}$Institute for Theoretical Physics, Utrecht University,
3508 TA Utrecht, The Netherlands}

\centerline{$^3$Martin-Luther-Universit\"at Halle-Wittenberg, 
Fachbereich Physik,
D-06099 Halle, Germany}}

\vskip .1in

\end{center}

\vskip .2in

\begin{center} {\bf ABSTRACT } \end{center}
We review modifications of the Bekenstein-Hawking area law for
black hole entropy in the presence of higher-derivative interactions. In   
four-dimensional $N=2$ compactifications of string theory or
M-theory these modifications are crucial for finding 
agreement between the macroscopic entropy obtained from
supergravity and the microscopic entropy obtained by counting
states in string or M-theory. Our discussion is based on the effective
Wilsonian action, which in the context of $N=2$ supersymmetric
theories is defined in terms of holomorphic quantities. At the end we briefly
indicate how to incorporate non-holomorphic corrections.  

\vfill

March 1999\\
\end{titlepage}

\section{Introduction}
It is one of the most intriguing properties of black holes in
general relativity that one can derive a set of laws, 
called the laws of black hole mechanics, which are formally equivalent
to the laws of thermodynamics \cite{Haw}.
For instance, the first law of thermodynamics, which relates variations of the
internal energy and the work done to the variation of the entropy, 
has its counterpart in black hole mechanics. The first law for black
holes in general relativity relates the variation of the mass and
angular momentum of the black 
hole to the change in the area of its event horizon. This then leads
to the celebrated Bekenstein-Hawking area law \cite{Bek}, which expresses the
black hole entropy in terms of the horizon area. 

The first law is quite a remarkable result as it relates global
quantities of the black hole, such as its energy or mass and its
angular momentum, which can be entirely determined from the behaviour
of the fields at spatial infinity, to the horizon area which is
defined at the inner boundary of the black hole solution. Also, 
the connection with thermodynamics suggests a possible 
interpretation of the entropy in terms of microstates. Such an 
interpretation has recently been provided in the context of string
theory \cite{StromVafa}. 

{From} the point of view of the field theory, be it fundamental or
effective, it is rather 
surprising that variations near the outer boundary at spatial infinity
are related to variations near the inner boundary at the
horizon. Moreover, an effective field theory action will contain more than
just the standard Einstein-Hilbert term and will depend on higher
derivatives of the fields. So one may wonder whether there could be a
natural principle that explains the behaviour implied by the
first law of black hole mechanics for more generic field
theories. Such a principle can be provided by making use of the concept of
a surface charge, and a specific proposal for this charge was put
forward by Wald \cite{Wald}. The surface 
charge is related to the conventional Noether current. In a certain
sense there is no Noether current associated with a local gauge
symmetry, but there exists a current associated with the residual invariance
of a background configuration. When evaluating this current subject to
the field equations, current conservation becomes trivial and the
current takes the form of an improvement term, i.e., the
derivative of an antisymmetric tensor. This antisymmetric tensor,
sometimes called the Noether potential, can be written down
for arbitrary gauge parameters. Integration of this potential over the
boundary of some (spatial) hypersurface leads to a surface charge,
which, when restricting 
the gauge transformation parameters to those that leave a certain
background invariant, is equal to the Noether charge in the usual
sense. Variations of this surface charge that continuously connect
solutions of the equations of motion thus relate the various surface
contributions and this is how eventually the first law of black hole
mechanics follows (this is, for example, reviewed in 
\cite{JacKaMy,IWald,WaldRev}). 

In this approach the entropy is related to the integral of the Noether
potential over the horizon. The entropy defined in this way is
a local geometric quantity and equals the Bekenstein-Hawking entropy
(i.e., one fourth of the horizon area) for Einstein gravity. 
But for more general actions containing terms of higher order in
derivatives the entropy formula contains additional terms. Since the
low-energy effective action constitutes the macroscopic 
or `thermodynamic' level of description of quantum gravity, the laws
of black hole mechanics have to hold for such effective actions
if they are more than accidental analogies. 

Naturally, since string theory is a
candidate for a theory of quantum gravity, one expects to be able to
describe black holes also 
at the microscopic or `statistical mechanics' level. Progress
into this direction has been made after nonperturbative 
dualities, D-branes and the M-theory description of 
string theory were discovered (see, for example, 
\cite{Pol,Mal} for a review). 
In particular, quantitative
agreement has been found between macroscopic black hole entropies extracted
from supergravity solutions and microscopic entropies calculated
by counting excitations of D-branes and M-branes. A crucial ingredient
in establishing agreement is extended supersymmetry, because one needs
to interpolate between 
different regimes of the theory, such as the low-energy regime and
the regime of small string coupling. This interpolation is possible
for extremal black holes which are BPS solitons of extended supergravity.

The resulting expressions for the
entropy concern extremal charged black holes and depend exclusively on
the electric and magnetic charges. These results are obtained in cases
where (some of) these charges are large, but sometimes it is also
possible to evaluate some of the subleading corrections. 
The comparisons 
between macroscopic and microscopic entropy have
mostly been made in situations where the distinction between the 
Bekenstein-Hawking area law and more general definitions of
macroscopic black hole
entropy are immaterial. Only recently has it become apparent that
microscopic entropy formulae contain subleading corrections which
on the macroscopic level are due to higher-derivative interactions in
the field theory \cite{MalStrWit,Vaf,CarDeWMoh}.
In this case the correct macroscopic definition of entropy is
crucial for finding agreement, as we will review below. 

The structure of this paper is as follows. First we will review the
derivation of the Noether potential for Yang-Mills theories and for
gravity in the context of field theories with higher derivatives. Then
we will discuss 
the laws of black hole mechanics and certain key notions of black hole
physics in the framework of general effective Lagrangians. 
Particular emphasis will be put on Wald's derivation of the first law
and the related definition of black hole entropy as a Noether charge.
Then we recall the special features of extremal black holes and
their relation to supersymmetry.  We briefly describe how special geometry
encodes the couplings of vector multiplets to $N=2$ supergravity
in the presence of a certain class of higher-order derivative terms
proportional to the square of the Weyl tensor. This part of the
discussion is based on the effective $N=2$ Wilsonian action, which is 
defined in terms of a holomorphic quantity.
Then we review our
work on the entropy of extremal $N=2$ black holes in the presence of
these higher-derivative curvature terms, which uses a definition of the
macroscopic entropy that deviates from the area law.
We conclude with a 
few examples of black hole solutions arising in string theory 
compactifications.  By
recalling results of microscopic entropy
calculations in the context of string and M-theory compactifications,
we show that they perfectly agree with macroscopic entropy
calculations based on Wald's definition of macroscopic entropy.
We also briefly discuss how to incorporate non-holomorphic
corrections to the effective Lagrangian into macroscopic
entropy formulae.

\section{Noether potential and charge}
For any local symmetry there exists a
globally-defined Noether potential, denoted here by an antisymmetric
tensor ${\cal Q}^{\m\n}(\phi, \xi)$, which is a local function of the fields
and of the gauge transformation parameters, here generically denoted
by $\phi$ and $\xi$ respectively \cite{Wald}. To derive this
potential, one starts from a gauge-invariant action and one derives
the Noether current as if one 
is dealing with a (rigid)  residual symmetry associated with a certain
background field configuration. To illustrate this we will first
present the case of a Yang-Mills theory as a pedagogical example, with
a gauge-invariant Lagrangian ${\cal L}(F_{\m\n},\nabla_{\!\rho} F_{\m\n}, \psi,
\nabla_{\!\m}\psi)$ depending on the field strenghts $F_{\m\n}$, matter 
fields $\psi$ and first derivatives thereof. We begin by multiplying the gauge
transformations with a test function $\epsilon(x)$. This test function
as well as its derivatives will satisfy certain boundary conditions,
such that the variation of the action is proportional to the field
equations, in 
accord with a modified (in view of the higher-derivative
interactions) version of Hamilton's 
variational principle. Hence we have 
\be
\begin{array}{rcl}
\delta A_\mu &=& \epsilon(x) \,\nabla_{\!\mu\,} \xi(x)\,,\\[1mm]
\d\psi &=& \epsilon \, \xi\, \psi\,, \end{array}
\qquad 
\begin{array}{rcl}
\delta F_{\mu\nu} &=&
2\, \partial_{[\mu}\epsilon \,\nabla_{\!\nu]}\xi +
\epsilon\,[\xi, F_{\mu\nu}]\, ,  \\[1mm]
\qquad \d \nabla_{\!\m} \psi &=& \pa_\m\epsilon \,\xi\, \psi +
\e \,\xi\,\nabla_{\!\m} \psi\,.  \end{array}
\eq
Here the gauge field, the field strength and the transformation
parameters $\xi$ are written as Lie-algebra valued quantities in the
representation relevant for $\psi$. The explicit variation of the action now
leads directly to the current (using that $\e$ and its first
derivative vanish at the boundary) and one finds (we suppress the
gauge-invariant inner product notation),
\be
J^\mu = 2\, {\cal L}^{\mu\nu}\,\nabla_{\!\nu\,}\xi - 2\nabla_{\!\rho\,} 
 {\cal L}^{(\rho,\mu)\nu}\, \nabla_{\!\nu\,} \xi 
+2 \,{\cal L}^{[\rho,\mu]\nu}\,\nabla_{\!\rho} \nabla_{\!\nu\,}\xi 
- {\cal L}^{\mu,\rho\sigma}\, [F_{\rho\sigma}, \xi] + {\cal L}^{\mu}\,
\xi\psi\,,
\label{currentg}
\eq
where\footnote{
  Note that the variations are defined such that we do not
  differentiate with respect to {\it independent} field
  components. Specifically, the variation of the Lagrangian takes the
  form $\d{\cal L} = {\cal L}^{\m\n}\d F_{\m\n} + \cdots$.
  } 
\be
{\cal L}^{\mu\nu}= {\pa{\cal L}\over \pa F_{\mu\nu}} \,,\qquad 
{\cal L}^{\rho, \mu\nu} = {\pa{\cal L}\over \pa \nabla_{\!\rho}
F_{\mu\nu}} \,, \qquad {\cal L}^{\mu} = {\pa {\cal L}\over \pa
\nabla_{\!\m} \psi}\,. 
\eq
Observe that the Bianchi identity implies ${\cal L}^{[\rho,\mu\nu]}=0$. 
The equations of motion for the gauge fields take the form 
\be
-2 \,\nabla_{\!\mu\,} {\cal L}^{\mu\nu}+2 \nabla_{\!\mu}
\nabla_{\!\rho\,} {\cal L}^{\rho, \mu\nu}+
 {\cal L}^{\nu,\rho\sigma} [F_{\rho\sigma},\cdot] -{\cal  L}^{\nu}\,\psi= 0
 \,. \label{gauge-field-equation}
\eq
We refrain from giving the matter field equation for $\psi$. By virtue of
the combined field equations one can verify that this current is
conserved. 

However, one can also use the field equation
\eqn{gauge-field-equation} directly on the current so as to obtain 
\be
J^\mu = \pa_\n \,{\cal Q}^{\m\n}\,,
\eq
where ${\cal Q}^{\m\n}$ is the Noether potential, which in the case at hand
takes the form
\be
{\cal Q}^{\m\n} = 
2 \,{\cal L}^{\mu\nu}\, \xi - 2\,\nabla_{\!\rho\,}{\cal
L}^{\rho, \mu\nu} \, \xi +  {\cal L}^{\rho,
\mu\nu}\, \nabla_{\!\rho\,} \xi  \,,
\label{chargeg}
\eq
and is thus a local function of the fields and of the transformation
parameter. 
Obviously the fact that the Noether current is conserved is now
trivial. Nevertheless the Noether potential is still tied directly to the
invariant action, up to terms that are exact, i.e., that can be written as
$\pa_\rho X^{\m\n\rho}$, with $X^{\m\n\rho}$ a totally antisymmetric
tensor, and up to improvement terms that vanish in the symmetric
background (see below).  
{From} the Noether potential one can determine the charge by
integrating over a closed  hypersurface of co-dimension two, in which case the
ambiguity drops out. For a spacelike hypersurface we then  determine
the Noether charge in the usual sense, but written as a surface
integral. This charge is associated with the residual gauge symmetry
corresponding to parameters that satisfy $\nabla_{\!\mu\,}\xi =0$ in the
background.  Here we observe that we have used a somewhat different
algorithm for computing (\ref{currentg}) than the more conventional
one used, for instance, in \cite{Wald,JacKaMy,IWald}.  Compared with
the latter, this reflects itself in the presence of an additional
improvement term in (\ref{currentg}) equal to $ \partial_{\nu} (
{\cal L}^{\rho,\mu\nu} \nabla_{\!\rho\,} \xi )$, and hence in the presence
of the last term in (\ref{chargeg}).  We thus see that both approaches
yield the same Noether potential up to a term which vanishes in the
symmetric background. 

Here we have assumed that the Lagrangian is gauge invariant, which
implies that 
the Noether current is proportional to the symmetry variations of the
fields (as one can for instance verify directly for
\eqn{currentg}). Consider now the situation where we are dealing with
a continuous variety of solutions of the field equations which are
left invariant under a corresponding continuous variety of residual
gauge transformations. The Noether current associated with these
residual transformations is then vanishing for every one of these
solutions, 
from which it follows that, under changes of the solution (and the
corresponding symmetry parameters $\xi$), the Noether potential must
vary into a term that is exact, i.e., 
\be
\delta Q^{\m\n} = \pa_{\rho\,}\omega^{\m\n\rho}\,,
\eq
where $\omega$ is a totally antisymmetric tensor. This implies that
the value of the 
integral of the Noether potential over the boundary of a hypersurface of
co-dimension one must be constant and equal to zero for all these
solutions, 
provided that the fields are regular on the hypersurface. 
While this example is therefore not so interesting, the situation
changes when the Lagrangian is only invariant modulo a total
derivative, because then the current does in general not vanish in 
symmetric background solutions. For Yang-Mills theory this happens, for 
example, when the Lagrangian contains Chern-Simons terms. 

For the case of gravity with general coordinate invariance, the
Lagrangian is never invariant under general coordinate
transformations, so here the current will not necessarily vanish in a
symmetric background. So let us follow the same procedure for gravity
and construct the Noether potential,
starting from an invariant action depending on the Riemann 
curvature, and on a matter field $\psi_{\m\n}$ (with no particular symmetry)
and its first derivative. After multiplying the diffeomorphisms with a
test function $\epsilon(x)$ with appropriate boundary 
conditions, the transformation rules read 
\beqa
\delta g_{\mu\nu}  &=& - \epsilon(x) \Big(\nabla_{\!\mu\,} \xi_\nu(x) +
\nabla_{\!\nu\,} \xi_\mu(x)\Big)  \,, \nonumber\\
\d \psi_{\m\n} &=& -\epsilon(x) \Big(\nabla_{\!\m\,} \xi^\sigma \,
\psi_{\sigma\n}
+ \nabla_{\!\n\,} \xi^\sigma \,\psi_{\m\sigma}  +
\xi^\sigma\,\nabla_{\!\sigma} \psi_{\m\n}\Big)   \,.  
\eeqa
We note the following useful equations for variations of connections and
curvatures,
\beqa
\delta\Gamma_{\nu\rho}{}^\sigma &=& {\textstyle{1\over 2}}
g^{\sigma\lambda} \Big[ \nabla_{\!\nu\,} \delta g_{\rho\lambda}  +
\nabla_{\!\rho\,} \delta 
g_{\nu\lambda} -  \nabla_{\!\lambda\,} \delta g_{\nu\rho}   \Big]\,, 
\nonumber\\
\delta R_{\mu\nu\rho\sigma} &=&  R_{\mu\nu[\rho}{}^\lambda \,\delta
g_{\sigma]\lambda} + 2\,\nabla_{[\mu} \nabla_{[\rho} \,\delta g_{\sigma]\nu]} 
\,.
\eeqa
The variation of the action, assuming the boundary conditions for
$\epsilon$, now yields the current
\beqa
J^\mu &=& \xi^\mu \,{\cal L} - 2 \,{\cal L}^{\mu\nu\rho\sigma}\Big[
R_{\lambda \nu\rho\sigma} \,\xi^\lambda + \nabla_{\!\nu} \nabla_{\!\rho\,}
\xi_\sigma \Big] 
 +4 \,\nabla_\rho  {\cal L}^{\mu\nu\rho\sigma}\, \nabla_{(\nu\,}\xi
_{\sigma)} \nonumber \\
&& - {\cal L}_\psi^{\m,\rho\sigma} \, \Big[ \nabla_{\!\rho\,}\xi^\lambda
\,\psi_{\lambda\sigma}  + 
\nabla_{\!\sigma\,} \xi^\lambda\, \psi_{\rho\lambda}  + \xi^\lambda
\,\nabla_{\!\lambda}  \psi_{\rho\sigma} \Big]  \nonumber\\
&& + \ft{1}{2} (\nabla_{\!\lambda\,} \xi_\rho +
\nabla_{\!\rho\,} \xi_\lambda)
\Big[ {\cal L}_\psi^{\m,\rho\sigma} \psi^{\lambda}{}_{\sigma}
+ {\cal L}_\psi^{\m,\sigma \rho} \psi_{\sigma}{}^{\lambda}
+ {\cal L}_\psi^{\rho,\m\sigma} \psi^{\lambda}{}_{\sigma} \nonumber\\
&& \hspace{35mm} +  {\cal L}_\psi^{\rho,\sigma \m} \psi_{\sigma}{}^{\lambda} 
 -  {\cal L}_\psi^{\rho,\lambda \sigma} \psi^{\mu}{}_{\sigma}
-  {\cal L}_\psi^{\rho,\sigma \lambda} \psi_{\sigma}{}^{\mu} \Big]
\,, \label{gravNcurrent}
\eeqa
where 
\be
{\cal L}^{\mu\nu}= {\pa{\cal L}\over \pa g_{\mu\nu}} \,,\qquad 
{\cal L}^{\mu\nu\rho\sigma} = {\pa{\cal L}\over \pa
R_{\mu\nu\rho\sigma}} \,, \qquad {\cal L}_\psi^{\rho,\mu\nu} = {\pa
{\cal L}\over \pa \nabla_{\!\rho} \psi_{\m\n}}\,. 
\eq
Observe that 
${\cal L}^{\mu\nu\rho\sigma}$ is antisymmetric in $[\m\n]$ and in
$[\rho\sigma]$; furthermore it is symmetric under pair exchange, ${\cal
L}^{\mu\nu\rho\sigma} ={\cal L}^{\rho\sigma\mu\nu}$, and satisfies the
cyclicity property ${\cal L}^{[\mu\nu\rho]\sigma}=0$. 

The current \eqn{gravNcurrent} is conserved by virtue of the equations
of motion for the metric,
\beqa
&&{\textstyle{1\over 2}} g^{\mu\nu} \,{\cal L} +  {\cal L}^{\mu\nu} +
{\cal L}^{\rho\sigma\lambda(\mu}\,R_{\rho\sigma\lambda}{}^{\nu)} 
- 2 \,\nabla_{\!(\rho} \nabla_{\!\sigma)} {\cal L}^{\rho\mu\nu\sigma}
\nonumber\\
&&+\ft14 \nabla_{\!\lambda}\Big[ {\cal L}_\psi^{\lambda,\mu\rho} \,
\psi^\n{}_\rho  +  
{\cal L}_\psi^{\lambda,\rho\mu} \,\psi_\rho{}^\n  +
{\cal L}_\psi^{\m,\lambda\rho} \,\psi^\n{}_\rho  \nonumber \\
&& \hspace{15mm}  +{\cal L}_\psi^{\m,\rho\lambda} \,\psi_\rho{}^\n  
-{\cal L}_\psi^{\m,\n\rho} \,\psi^\lambda{}_\rho  -{\cal
L}_\psi^{\m,\rho\n} 
\,\psi_\rho{}^\lambda   + (\m\leftrightarrow\nu)\Big] =0   \,,
\label{metric-eq-motion} 
\eeqa
and of the equation of motion for $\psi_{\m\n}$, 
\be
\nabla_{\!\rho\,}{\cal L}_\psi^{\rho,\m\n} - {\cal L}_\psi^{\m\n} =
0\,,  \label{psi-eq-motion}
\eq
where
\be 
{\cal L}^{\m\n}_\psi = {\pa{\cal L}\over \pa \psi_{\m\n}}\,,
\eq
as well as of the general covariance of the Lagrangian. The latter
implies
\beqa
&&2\,{\cal L}^{\mu\nu} \,\nabla_{\!\mu\,} \xi_\nu  +4\, {\cal
L}^{\mu\nu\rho\sigma}\,R_{\mu\nu\rho}{}^\lambda \,
\nabla_{\!\sigma\,}\xi_{\lambda} + {\cal L}_\psi^{\m\n}(\nabla_{\!\m\,}
\xi^\rho \,\psi_{\rho\n}
+\nabla_{\!\n\,}\xi^\rho \,\psi_{\m\rho} ) \nonumber\\
&&+ {\cal L}_\psi^{\rho,\m\n} (\nabla_{\!\rho\,}\xi^\sigma
\,\nabla_{\!\sigma\,}\psi_{\m\n} + \nabla_{\!\m\,}\xi^\sigma 
\,\nabla_{\!\rho\,}\psi_{\sigma\n}+  \nabla_{\!\n\,}\xi^\sigma
\,\nabla_{\!\rho\,} \psi_{\m\sigma}  ) =0  \,,\;
\eeqa
which in turn gives rise to the identity
\beqa
&&2{\cal L}^{\mu\nu} \,\xi_\nu  -4\, {\cal
L}^{\mu\nu\rho\sigma}\,R_{\rho\sigma\nu}{}^{\lambda} \,
\xi_{\lambda} 
+{\cal L}_\psi^{\m\n}\xi^\sigma \,\psi_{\sigma\n} + 
{\cal L}_\psi^{\n\m}\xi^\sigma \,\psi_{\n\sigma}  \nonumber\\
&& 
+ {\cal L}_\psi^{\m,\n\rho} \xi^\sigma
\,\nabla_{\!\sigma\,}\psi_{\n\rho} 
+ {\cal L}_\psi^{\rho,\m\n} \xi^\sigma
\,\nabla_{\!\rho\,}\psi_{\sigma\n}  
+ {\cal L}_\psi^{\rho,\n\m} \xi^\sigma
\,\nabla_{\!\rho\,}\psi_{\n\sigma} 
=0  \,. \label{covariance}
\eeqa
Imposing the equations of motion \eqn{metric-eq-motion} and
\eqn{psi-eq-motion} and the relation
\eqn{covariance}, the current takes the form $J^\m= \nabla_\n {\cal
Q}^{\m\n}$ with the Noether potential equal to 
\beqa
{\cal Q}^{\mu\n} &=& 
- 2\, {\cal L}^{\mu\nu\rho\sigma} \,\nabla_{\!\rho\,}\xi_\sigma 
+ 4 \,\nabla_{\!\rho\,}{\cal L}^{\mu\nu\rho\sigma} \, \xi_\sigma \nonumber\\
&& +\ft12 \Big[-{\cal L}_\psi^{\m,\n\rho} \,\psi^\sigma{}_{\rho} 
-{\cal L}_\psi^{\m,\rho\n} \,\psi_{\rho}{}^\sigma 
+{\cal L}_\psi^{\m,\sigma\rho} \,\psi^\n{}_{\rho} \nonumber \\
&&\hspace{9mm} 
+{\cal L}_\psi^{\sigma,\m\rho} \,\psi^\n{}_{\rho}
+{\cal L}_\psi^{\m,\rho\sigma} \,\psi_{\rho}{}^\n 
+{\cal L}_\psi^{\sigma,\rho\m} \,\psi_{\rho}{}^\n -
(\mu\leftrightarrow\n)  
 \Big] \,\xi_\sigma  \,.
\label{noetgrav}
\eeqa
When spacetime exhibits an isometry with a corresponding Killing
vector $\xi^{\mu}$, so that 
\be
\nabla_{\!\mu\,} \xi_{\nu} + \nabla_{\!\nu\,}  \xi_{\mu} = 0\,,
\label{Killing}
\eq
the corresponding Noether potential is only proportional
to $\xi^\m$ and to 
its curl $\nabla_{[\mu\,}\xi_{\n]}$ in view of the identity
\be
\nabla_{\!\mu} \nabla_{\!\nu\,} \xi_{\rho} = 
 R_{\nu \rho \mu}{}^\sigma \xi_{\sigma}\,.
\label{KillingA}
\eq
The Noether charge associated with the isometry  is given
as the integral of the Noether potential over the boundary of a
spatial hypersurface.  The variation of the charge under 
infinitesimal variations of the background solution will be discussed in
the following section.

\section{Black hole entropy}

As was shown by Wald one can employ the Noether charge in order to find
generalized definitions of the black hole entropy that are consistent
with the first law of black hole mechanics. Although most of the
following applies to black holes in any number of spacetime dimensions
we will restrict ourselves to four dimensions. As we discussed already in
the introduction, the description in terms of a surface charge
can in principle explain how variations of the fields subject to the
equations of motion at the horizon can be expressed in terms of the
variations of quantities that are defined at spatial infinity. The
crucial observation \cite{Wald} is that there exists a Hamiltonian,
whose change under a variation of the fields can be expressed in terms
of the corresponding change of the Noether charge and takes the
following form 
\be
\d H = \d\Big (\int_C d\Omega_\m \;J^\mu  \Big )  - \int_C d\Omega_\m
\: \nabla_{\!\n\,} ( \xi^\m\,\theta^\n - \xi^\n \,\theta^\m) \,,
\label{variation1} 
\eq
where we integrate over a Cauchy surface $C$ with volume element
$d\Omega_\m$,  
$J^\m$ is the Noether current associated with a particular
Killing vector $\xi^\m$ (to be discussed below) and $\theta^\m$ is
defined by the surface integral that one 
obtains when considering the change  of the action under the 
field variation,  
\be
\d S = \int d^4x\; \pa_\m \Big(\sqrt{-g} \,\theta^\m \Big)\,,
\eq
after subsequently imposing the field equations. 
We assume that the
Killing vector field is timelike so that $H$ can be associated 
with a Hamiltonian that governs the evolution along the integral
timelike lines of $\xi^\m$. 

For black holes, $\xi^\m$ is a horizon-generating Killing field, 
meaning that it is the normal vector of a null hypersurface,
called the Killing horizon. The Cauchy surface in \eqn{variation1} is
chosen to extend from spatial infinity down to the 
Killing horizon where the Killing field turns lightlike.  In Einstein
gravity it can be shown that
under certain assumptions, such as that the dominant-energy
condition holds and that the matter field equations of motion have a
well defined Cauchy problem, all event
horizons are Killing horizons \cite{HawEll}, but in more general theories
this is not obvious. For static black holes, 
the event horizon is always a Killing horizon
\cite{Carter}, irrespective of the precise form of the action. In that
case  the  relevant  
Killing field is just the static Killing vector field. In the
following we shall always assume that we are dealing with a Killing
horizon. 

When imposing the equations of motion, \eqn{variation1} takes the form of
surface integrals over the boundary $\pa C$ with surface element
$d\Sigma_{\m\n}$,
\be 
\d H= \int_{\pa C} d\Sigma_{\m\n} \, \Big(\delta 
{\cal Q}^{\m\n} -  \xi^\m\,\theta^\n +
\xi^\n \,\theta^\m\Big) \,.
\label{deltah}
\eq
Here it is important that the last two terms are proportional to the
Killing vector and not to its curl. Furthermore one assumes that these 
terms can be rewritten as variations of some (not necessarily globally
defined) quantity. Because one can prove that  $\d H=0$ (but not that
$H$ itself vanishes) whenever
$\xi^\m$ is a Killing vector characterizing an invariance of the
full background, it follows that the 
sum of the surface integrals in \eqn{deltah} has to vanish! If one
identifies, up to a certain proportionality factor, the resulting
variations of the surface 
integrals at infinity with the mass and angular 
momentum variations that one has in the first law, then the surface
integral at the horizon defines the variation of the entropy. Hence
the mass, the angular momentum and the entropy are all surface charges
derived from the same current, and the entropy takes the form of an
integral over the Noether potential \cite{Wald},  
\be
{\cal S} =  -\pi \, \int_{\Sigma_{\rm hor}} {\cal Q}^{\mu \nu} \;
\epsilon_{\mu \nu}\; \Big\vert_{\xi^\m=0,\; \nabla_{[\m\,}\xi_{\n]}=
\e_{\m\n}}\;.  
\label{entronoet}
\eq
Here $\Sigma_{\rm hor}$ denotes a spacelike cross section of the
Killing horizon (which usually has the topology of $S^2$) and we
have used $d\Sigma_{\m\n} = 
\e_{\m\n}\,\sqrt{h} \,d^2x$. Here 
$\epsilon_{\mu \nu}$ denotes the binormal, whose definition we will
review below and which 
is normalized according to $\e_{\m\n}\e^{\m\n}=-2$; $\sqrt{h}
\,d^2x$ gives the surface element induced on $\Sigma_{\rm hor}$. We already
mentioned that the Noether potential ${\cal Q}^{\m\n}$ can be
decomposed according to 
${\cal Q}^{\mu \nu} = Y^{\mu \nu \rho \sigma}
\nabla_{[\rho\,}\xi_{\sigma]}+N^{\mu \nu \rho} \, \xi_{\rho} $. 
We shall discuss in due course how one is led to impose the conditions
$\xi^\m=0$ and $\nabla_{[\m\,}\xi_{\n]}=
\e_{\m\n}$ on the Killing vector field at $\Sigma_{\rm hor}$
associated with the Noether potential in \eqn{entronoet}. 
Let us already point out that the
contributions we are suppressing in (\ref{entronoet})
by imposing these conditions have actually 
been shown to vanish for non-extremal
black holes \cite{JacKaMy};
this is more subtle for extremal black holes,
where the surface gravity is vanishing.  Nevertheless we simply
adopt \eqn{entronoet} as the definition for the entropy in both the
extremal and non-extremal case. Note that in both cases
\eqn{entronoet} leads in principle to a well-defined result. 

Finally, the normalization in \eqn{entronoet} has been chosen such
that we reproduce the Bekenstein-Hawking area
law for static black holes in general relativity. 
To see this, we note that the Lagrangian associated to Einstein
gravity, with the conventions of \cite{CarDeWMoh}, 
reads $8 \pi {\cal L} =  - \ft{1}{2} R$, so that $8\pi\,{\cal
L}^{\m\n\rho\sigma}= -\ft12 g^{\m[\rho}\,g^{\sigma]\n}$. Consequently we
have, subject to the conditions in  \eqn{entronoet}, that ${\cal
Q}^{\mu \nu} =  - 2  \, 
{\cal L}^{\mu\nu\rho\sigma} \, \epsilon_{\rho \sigma}$ which equals
$(8\pi)^{-1}\,\e^{\m\n}$ for Einstein gravity. Substitution of this
result into \eqn{entronoet} immediately yields one-fourth of the area,
expressed in Planck units. 

In the presence of higher-derivative curvature terms, the entropy
of a static black hole solution will in general not any longer be simply
given by the Bekenstein-Hawking area law.  As shown in the previous
section (c.f. \eqn{noetgrav}), when the Lagrangian
depends on the Riemann curvature tensor but not on derivatives
thereof, as well as on matter fields with at most second
derivatives, then ${\cal Q}^{\mu \nu} = - 2  \,
{\cal L}^{\mu\nu\rho\sigma} \, \epsilon_{\rho \sigma}$ when we impose
the conditions on the Killing vector at the horizon, specified in
\eqn{entronoet}. The entropy of the static black hole is then given by
\cite{Visser,JacKaMy,IWald}
\be
{\cal S} = 2\pi \int_{\Sigma_{\rm hor}} \;  {\cal L}^{\mu \nu \rho
\sigma} \; \epsilon_{\mu \nu}  \;  \epsilon_{\rho  \sigma} \;.
\label{graventro}
\eq
This is the result that we will need in the next section. 

We will now review some of the concepts involved in the definition
(\ref{entronoet}) of the black hole 
entropy \cite{Wald,JacKaMy,IWald,Waldbook} 
in somewhat more detail.
Wald's construction of the entropy as a surface charge applies to stationary
and to some extent also to non-stationary black holes in arbitrary
spacetime dimensions, but in the following we will for concreteness
restrict the discussion
to four-dimensional static black hole solutions.
A spacetime containing a black hole consists
of an asymptotically flat region, which is separated by a (future) event 
horizon from an interior region, such that after crossing the 
horizon one is trapped in the interior region.  
Usually, the term `horizon' designates either
the `horizon at a given instant of time', i.e. 
a spacelike two-surface $\Sigma_{\rm hor}$,  or the corresponding
worldvolume $\Delta$ swept out in time by $\Sigma_{\rm hor}$, 
which is a hypersurface in spacetime.  In order to avoid any confusion, 
we will consistently distinguish between these two cases and 
use the symbols $\Sigma_{\rm hor}$ and $\Delta$ throughout.

A hypersurface can be defined by an equation $f(x^{\mu})=0$.
Since the gradient $\nabla_{\mu} f$ is automatically
normal to the surface (i.e. $t^{\mu}\, \nabla_{\mu} f = 0$ for all
tangent vectors $t^{\mu}$) one can equally well specify 
this surface in terms of
the normal vector field $n_{\mu} = \nabla_{\mu} f$. 
According to the 
Frobenius theorem, which formulates necessary and sufficient conditions
for vector fields to define smoothly embedded submanifolds, 
a vector field
$n^{\mu}$ is hypersurface orthogonal in the above sense
if and only if
\be
n_{[\mu} \nabla_{\nu} n_{\rho]} = 0 \;.
\label{Frobenius}
\eq
If the normal vector field of a hypersurface is a 
null vector field, 
$n_{\mu} n^{\mu} =0$, then the hypersurface is called a null hypersurface.
Note that the 
normal vector field of a null hypersurface is 
also tangential to it. The (future) event horizon $\Delta$ is such a
null hypersurface.  Since we assumed that it is a Killing horizon we
know that its normal vector is in fact a Killing vector field.  

By taking a spacelike cross section of $\Delta$ one obtains 
a spacelike surface $\Sigma_{\rm hor}$.
In the two-dimensional space normal to $\Sigma_{\rm hor}$ there exists
one linearly independent 
antisymmetric tensor of second rank, $\epsilon_{\mu\nu}$, which is called
the normal bivector or simply the binormal.  We will normalize it
according to
$\epsilon_{\mu \nu} \, \epsilon^{\mu \nu} = -2$. 
The binormal can be explicitly written as a bivector, as 
follows.  First we note that the space normal to $\Sigma_{\rm hor}$
has signature $(-+)$, and therefore it can be spanned by
two null vectors. Since $\Sigma_{\rm hor}$ is contained in $\Delta$,
one of the null vectors can be taken to be the normal 
$n^{\mu}$ of $\Delta$, that is, one of the null vectors is proportional to 
the Killing vector $\xi^{\mu}$.  The other null vector, which we denote
by $N^{\mu}$, is chosen such that
$N^{\mu} n_{\mu} = -1$.
Then, the bivector
\be
\epsilon_{\mu \nu} = N_{\mu} n_{\nu} - N_{\nu} n_{\mu}\,
\label{bivector}
\eq
is non-vanishing in the normal directions and has the
required normalization.

The spacetime metric describing a static and spherically symmetric
black hole solution can be written as
\be
ds^2 = - e^{2g(r)}dt^2 + e^{2 f(r)} ( dr^2 + r^2  d \Omega^2 )
\label{StaticMetric}
\eq
in isotropic coordinates $(t,r,\phi,\theta)$.
In such an adapted coordinate system, the time coordinate
$t$ and the radius $r$ denote the
directions normal to the two-sphere $\Sigma_{\rm hor}$, whose coordinates are
$\phi$ and $\theta$.  Thus, the
associated  binormal $\epsilon_{\mu \nu}$ has the 
non-vanishing components (up to an overall sign) 
$\epsilon_{tr} = - \epsilon_{rt}= \exp[ g(r)+f(r)]$.   

In order to formulate the laws of black hole mechanics, we need
to define the notion of 
surface gravity $\kappa_{\rm S}$ associated with a Killing horizon.
Since the Killing vector field is null on the horizon (but in general
$\xi^\m\not=0$), we can take $f=\xi^{\mu} \xi_{\mu} = 0$ as the defining
equation of the horizon $\Delta$. 
Since $\nabla_{\mu} f$ is normal to $\Delta$, it must be 
proportional to $\xi_{\mu}$ itself. The coefficient of proportionality
defines the surface gravity $\kappa_{\rm S}$ of the black hole:
\be
\nabla_{\!\mu\,}(\xi^{\nu} \xi_{\nu}) = -2 \kappa_{\rm S}\, \xi_{\mu}\,.
\label{kappa}
\eq
Observe that this presupposes a certain intrinsic normalization for the
Killing vector field, which is usually specified at spatial infinity. 
Subsequently one shows that
\be
\kappa^{2}_{\rm S} = - \ft{1}{2} (\nabla^{\mu}  \xi^{\nu})( \nabla_{\mu}
\xi_{\nu}) = \ft{1}{2} R_{\mu \nu} \, \xi^{\mu} \xi^{\nu} \;\;\;\; 
{\rm and} \;\;\;\; \xi^{\mu}\, \partial_{\mu} \kappa_{\rm S} = 0
\,.
\label{kappasquared}
\eq 
The first equation follows by multiplying
\eqn{kappa} with $\nabla^\m\xi^\rho$ and by making use of the
Killing equation (\ref{Killing}) 
together with the Frobenius theorem (\ref{Frobenius}) with
$n^\m=\xi^\m$.  On the other hand, 
it follows from \eqn{KillingA} that
$\xi^\rho\,\nabla_{\!\rho\,} \nabla_{\!\m\,}\xi_\n=0$, so that
$\kappa_{\rm S}$ is constant along the integral curves 
of $\xi^{\mu}$ on $\Delta$.  This is the last equation.
The second equality is then obtained
by applying a covariant derivative $\nabla^{\mu}$ on (\ref{kappa}) and by 
using (\ref{KillingA}).
For a static and spherically symmetric black hole, 
the surface gravity $\kappa_{\rm S}$ is thus constant over all of $\Delta$.
The constancy of $\kappa_{\rm S}$ over $\Delta$ is known as the 
zero-th law of black hole mechanics. We should point out here that the
above considerations do not quite apply to extremal black holes,
because they have zero surface gravity. We return to this point
shortly.

The comparison with the laws of thermodynamics suggests to identify
the surface gravity with the temperature of the black hole,
up to a multiplicative constant. Then, the zero-th law is reinterpreted by 
stating that the Killing horizon is in thermodynamical equilibrium
and that it radiates like a black body. This interpretation is confirmed
by the phenomenon of Hawking radiation which is found when
quantizing matter fields in a classical black hole background \cite{Bek}. 
In this way the proportionality constant is fixed according to
$T = {\kappa_{\rm S}}/{2 \pi}$, where $T$ denotes the Hawking
temperature.

It is instructive to calculate the surface gravity for a 
Reissner-Nordstr{\o}m black hole in general relativity, which provides
an example of a static and charged black hole.  In a coordinate system where
the curvature singularity is located at $r=0$, 
the associated spherically symmetric spacetime
line element is given by
\be
ds^2 = - e^{2h(r)} dt^2 + e^{-2h(r)} dr^2 + r^2 d \Omega^2 \;\;,\;\;
e^{2h(r)} = 1 - \frac{2M}{r} + \frac{Q^2}{r^2} \;,
\label{rn}
\eq
where $M$ and $Q$ denote the mass and the charge of the black hole, 
respectively. 
The cases $Q=0$, $M > |Q|$ and $M=|Q|$ yield the Schwarzschild,
the non-extremal and the extremal Reissner-Nordstr{\o}m 
black hole, respectively.
The case $M<|Q|$ is excluded by the generalized positivity theorem
for the ADM mass when assuming the so-called dominant-energy condition,
i.e., a non-negative energy density and a non-spacelike energy flow
for every observer \cite{GibHul}.
The surface gravity is computed to be \cite{Waldbook}
\be
\kappa_{\rm S} = 
\frac{ \sqrt{M^{2}-Q^{2}}}{2M(M + \sqrt{M^{2}-Q^{2}}) -Q^{2}} \;\;,
\eq
which shows that charged and neutral black holes behave very
differently when loosing energy by Hawking radiation. The Schwarzschild
black hole ($Q=0$) will heat up in the process, because
$\kappa_{\rm S} = (4M)^{-1}$, which suggests
that it will completely evaporate into radiation. On the 
other hand  a charged black hole will cool down when approaching the
extremal limit, namely $\kappa_{\rm S} \rightarrow 0 $ for $M \rightarrow Q$.
The final state, an extremal black hole with $\kappa_{\rm S} =0$, has
vanishing Hawking temperature and could therefore be a stable
object.
In the following, we will always call black hole solutions 
extremal if they have $\kappa_{\rm S}=0$.
Note that this does not necessarily 
imply that they are extremal in the sense of being
on the edge of developing a naked singularity.

We now turn to the
first law of black hole mechanics which relates changes
of the black hole mass to changes in the entropy as well as to changes
of other quantities which characterize the black hole, such as its charges.
In the following we will omit these latter changes for
the sake of clarity.
For a static black hole, the first law 
follows from (\ref{deltah}) with $\d H=0$ by taking  
the Cauchy surface $C$ to have a boundary $\Sigma_{\rm hor} \cup
\Sigma_{\infty}$,  where $\Sigma_{\rm hor}$
denotes a spacelike cross section of the event horizon, and where 
$\Sigma_{\infty}$
denotes a two-sphere at spatial infinity.  Inspection of (\ref{deltah})
and \eqn{entronoet} (where in (\ref{entronoet}) 
we have adopted a different normalization for the
Killing vector, see below)
then suggests to identify
the change in the mass and in the entropy of the black hole with
\beqa 
\d M&=& - \ft12 \int_{\Sigma_{\infty}} 
d\Sigma_{\m\n} \, \Big( \delta {\cal Q}^{\m\n} -  \xi^\m\,\theta^\n +
\xi^\n \,\theta^\m\Big) \;,\nonumber\\
\frac{\kappa_{\rm S}}{2\pi} \,
\d {\cal S}&=& -\ft12 \int_{\Sigma_{\rm hor}} d\Sigma_{\m\n} \,
\Big(\delta {\cal Q}^{\m\n} -  \xi^\m\,\theta^\n + \xi^\n
\,\theta^\m\Big) \,,\; 
\label{first}
\eeqa
in accordance with the first law of thermodynamics,
$\delta E = T \delta {\cal S} + \cdots$, with $E = M$ and
$T= \kappa_{\rm S}/{2 \pi}$.  We  note that the first law applies to
non-extremal black holes, which have a non-vanishing surface
gravity. We now proceed to express the entropy as a surface integral
of a  
local geometrical quantity.

As already mentioned, the Noether potential has the generic form
${\cal Q}^{\mu \nu} =  Y^{\mu \nu \rho \sigma}
 \nabla_{[\rho\,}\xi_{\sigma]} + N^{\mu \nu \rho} \xi_{\rho}$, where
the tensors $N^{\mu \nu \rho}$ 
and $Y^{\mu \nu \rho \sigma}$ are local quantities constructed out of
the Riemann tensor and its derivatives.  
The antisymmetric tensor $\nabla_{\!\mu\,} \xi_{\nu}$ can be decomposed as
follows,
\be
\nabla_{\!\mu\,} \xi_{\nu} 
=  \kappa_{\rm S} \, \epsilon_{\mu \nu} + t_{[\mu} \,\xi_{\nu]} \,,
\label{decomp}
\eq
where $t_{\mu}$ is tangential to $\Sigma$.  This decomposition expresses
the fact that according to the Frobenius theorem  
the non-vanishing components of $\nabla_{\mu} \xi_{\nu}$
are those where at least one
of the indices is not tangential to $\Sigma$. The coefficient
of $\epsilon_{\mu \nu}$ is determined by
contracting (\ref{decomp})
with $\nabla^{\mu} \xi^{\nu}$ and by comparing with
(\ref{kappasquared}), where one also uses the explicit realization
(\ref{bivector})
of $\epsilon_{\mu \nu}$ as a bivector.
By substituting the decomposition (\ref{decomp}) into ${\cal Q}^{\mu \nu}$
we obtain ${\cal Q}^{\mu \nu} = \kappa_{\rm S} \, 
Y^{\mu \nu \rho \sigma}  \epsilon_{\rho\,\sigma} + [ N^{\mu \nu \lambda}  + 
Y^{\mu \nu \rho \sigma}  \, t_{[\rho}\,\d^\lambda{}_{\!\sigma]\,}]\,
\xi_{\lambda}$. Observe that the last two terms of the integrand in
\eqn{first} are proportional to the Killing vector and can thus be
absorbed into $N^{\m\n\rho}$, so they don't need to be discussed
separately.  

For non-extremal black holes there is a theorem \cite{RacWal}
which states that the horizon hypersurface
$\Delta$ contains, or can be analytically extended to contain,
a spacelike cross section, called the
bifurcation surface $\Sigma_{0}$,
where the timelike Killing
vector field has a zero,
$\xi^{\mu} = 0$, so that $\nabla_{\mu} \xi_{\nu} = 
\kappa_{\rm S} \epsilon_{\mu \nu}$.  In theories with matter one also has
to assume that the matter fields can likewise be analytically continued
and that all the fields are regular at the bifurcation surface.
Thus, by evaluating the Noether potential on
$\Sigma_{0}$ one can get rid of the terms in ${\cal Q}^{\mu \nu}$
proportional to the Killing vector field, so that we are left with
$\kappa_{\rm S}/2\pi$ times the variation of the entropy defined in
\eqn{entronoet} but with $\Sigma_{\rm hor}$ replaced by $\Sigma_0$. 
As already mentioned, it can be shown \cite{JacKaMy} that, when replacing 
$\Sigma_{0}$
with any other spacelike cross section $\Sigma_{\rm hor}$ 
of the horizon, the resulting 
expression for the entropy is given by a similar
expression, namely by (\ref{entronoet}), which
indeed expresses the entropy as a surface integral of a local geometrical
quantity over an arbitrary cross section of the horizon.


As stressed above, the above procedure for deriving the first law of
black hole 
mechanics applies to non-extremal black holes.  Nevertheless,
the resulting expression for the entropy (\ref{entronoet})
remains well behaved in the extremal limit $\kappa_{\rm S} \rightarrow 0$,
as it is independent of $\kappa_{\rm S}$ and of the Killing field $\xi^{\mu}$.
Since extremal black holes do not possess a
bifurcation surface, it is important that the entropy can be evaluated 
on any spacelike cross section $\Sigma_{\rm hor}$, as is the case with 
(\ref{entronoet}).
Thus, we expect that the entropy of an extremal
black hole, if computed from (\ref{entronoet}), will be non-vanishing in
general.  It should be pointed out though that the question whether or not
an extremal black hole has a non-vanishing entropy, is a somewhat
subtle issue that depends on the approach used to compute
its entropy.  For instance, 
in the context of semiclassical quantization
of matter coupled to Einstein gravity, 
one finds that the entropy of an extremal Reissner-Nordstr{\o}m
black hole is non-vanishing and given by 
the Bekenstein-Hawking area law, provided the extremal limit is taken
after quantization.  If, on the other hand, the quantization is performed
after extremalization (that is, if the
quantization is applied to the part of
phase space that only contains extremal configurations), 
then the resulting entropy is vanishing.
This has been shown both in the Euclidean path integral framework
\cite{GhoMit} and in the Minkowskian canonical framework \cite{KieLou}.
In the context of string theory, various
microscopic state countings
yield a non-vanishing entropy
in all cases where extremal black holes have a non-vanishing horizon
area. Thus, string theory favours to treat extremal black holes
as limiting cases of non-extremal ones.

A comparison of Wald's Noether charge approach with various other
approaches can be, for instance, found in \cite{Visser,IyeWal2} and 
references therein.  Within their domain of applicability, all of
these other approaches yield results in agreement with the Noether
charge approach.

\section{Supersymmetric black holes}

The electrically charged static
extremal Reissner-Nordstr{\o}m solution of Einstein-Maxwell 
theory, which we briefly described in (\ref{rn}), 
enjoys several remarkable properties, as follows.
It interpolates between two maximally symmetric spaces, namely
Minkowski flat spacetime at spatial infinity and 
Bertotti-Robinson spacetime at the
horizon. There exists a dyonic version of it, whose mass $M$ and
entropy ${\cal S}$ (the latter computed from the area law) are
determined in terms of its electric and magnetic charges $q$ and $p$
as $M = |q + ip |$  
and ${\cal S}= \pi |q + i p|^2$.  Since its temperature vanishes, this
suggests that such a dyonic black hole is quantum mechanically stable. 
Moreover,  its mass saturates the Bogomolnyi bound which follows
from the generalized positivity theorem for the
ADM mass \cite{GibHul}.  There also exists a static multi-center
version of it, which is described 
by the Majumdar-Papapetrou metric \cite{MajPap}
and which resembles the static multi-monopole 
solutions of Yang-Mills-Higgs theories in the Prasad-Sommerfield limit.

All these properties can be explained in terms of a symmetry principle,
namely supersymmetry, by embedding Einstein-Maxwell theory into
$N=2$ supergravity \cite{GibHul}. Then the extremal Reissner-Nordstr{\o}m
black hole solution 
can be interpreted as a supersymmetric soliton which interpolates
between two maximally supersymmetric
vacua of $N=2$ supergravity.  Globally the solution is invariant under
4 of the 8 supersymmetries. 
Its mass formula follows from the $N=2$ supersymmetry algebra and
takes the form $M=|z|$, where $z$ denotes the central charge of the
supersymmetry algebra.  It is determined in terms of the electric and 
magnetic charges $q$ and $p$ associated with the 
gauge field, namely by $z=q+ip$.
 
More recently, following the work of \cite{FerKalStr}, this has
been extended to the study of static supersymmetric black hole solutions
in four-dimensional 
theories describing the coupling of $n$ abelian vector multiplets
to $N=2$ supergravity in the presence of a certain class of $R^2$-terms
\cite{BCDWLMS,CarDeWMoh}. Such theories arise as 
low-energy effective field theories of string and
M-theory compactifications on suitable compact
manifolds.  These effective Lagrangians
contain, in general, various other terms describing the couplings of additional
sectors, such as matter associated with hypermultiplets, to supergravity.
These other sectors, however, play only a limited role in the
following and we will therefore omit them.

Let us first review how the couplings of vector multiplets to $N=2$
supergravity can be described in terms of special geometry.  
Since we will allow for the presence of certain $R^2$-terms,
we will utilize the so-called superconformal framework, which provides
a systematic and powerful approach for constructing these couplings
\cite{deWvPr}.  It makes use of the fact that a conformal theory
of $N=2$ supergravity with suitable couplings to compensating
multiplets is gauge equivalent to $N=2$ Poincar\'e supergravity.
In the superconformal framework, there is a multiplet, the so-called
Weyl multiplet, which comprises the gravitational degrees of freedom, namely
the graviton, two gravitini as well as various other superconformal
gauge fields and also some auxiliary fields.  One of these
auxiliary fields is an anti-selfdual Lorentz tensor field $T^{ab\, i j}$,
where $i,j=1,2$ denote chiral $SU(2)$ indices (we recall that the 
associated $SU(2)$ algebra is part of the $N=2$ superconformal algebra).
The field strengths corresponding to the various gauge fields in the Weyl
multiplet reside in a so-called reduced chiral multiplet, denoted
by $W^{ab\, i j }$, from which one then constructs the unreduced
chiral multiplet $W^2 = (W^{ab\,i j } \varepsilon_{ij})^2$
\cite{BDRDW}.
The lowest component field of $W^2$ is ${\hat A} = 
(T^{ab\, i j } \varepsilon_{ij})^2$.

In addition, there are $n + 1$ abelian
vector multiplets labelled by an index $I = 0,
\dots, n$.  Each vector multiplet
contains a complex scalar field $X^I$, a vector gauge field $W_{\mu}^I$
with field strength $F_{\mu \nu}^I$, as well as two gaugini and a set
of auxiliary scalar fields.  The couplings of these $n+1$ vector multiplets
to the Weyl multiplet are encoded in a holomorphic function $F(X^I,
{\hat A})$, which is homogenous of degree two and consequently
satisfies $X^I F_I + 2 {\hat A} F_{\hat A} = 2 F$, where $F_I = 
\partial F/\pa {X^I}$, $F_{\hat A} = \partial F/\pa {\hat A}$.

The field equations of the vector multiplets are subject to 
equivalence transformations corresponding to electric-magnetic
duality, which do not involve the fields of the Weyl multiplet.
These equivalence transformations are symplectic 
${\rm SP}(2n+2;{\rm \bf Z})$
transformations.  Two complex $(2n+2)$-component vectors can now be
defined which transform linearly under 
${\rm SP}(2n+2;{\rm \bf Z})$ transformations,
namely
\be
V =  \left( \begin{array}{c} X^I \\ F_I(X,{\hat A}) \\ \end{array} \right)
\;\;\;\;\; {\rm and} \;\;\;\;\;
 \left( \begin{array}{c} F_{\mu \nu}^{\pm I} \\ 
G^\pm_{\mu \nu I}  \\ \end{array} \right)\,,
\eq
where $(F_{\mu \nu}^{\pm I},G^\pm_{\mu \nu I})$ denotes the
(anti-)selfdual part 
of $(F_{\mu \nu}^{I},G_{\mu \nu I})$.  The field strength $G^\pm_{\mu \nu I}$
is defined by the variation of the action with respect to 
$F^{\pm I}_{\mu \nu}$. The precise definition reads 
$G^{\pm}_{\mu \nu I} = 
- 4 i \,(-g)^{-1/2}\, \pa S / \pa F^{\pm I}_{\mu \nu}$.
By integrating the gauge fields over two-dimensional
surfaces enclosing their sources, it is possible to associate
to $(F_{\mu \nu}^{\pm I},G^{\pm}_{\mu \nu I})$ a symplectic vector $(p^I,q_I)$
comprising the 
magnetic and electric
charges.  It is then possible to construct a complex quantity $Z$
out of the charges and out of $V$ which is invariant under symplectic 
transformations, as follows, 
\be
Z =  e^{ {\cal K}/2} \, (p^I F_I (X,{\hat A}) - q_I X^I) \,,
\label{z}
\eq
where $e^{-\cal K} = i [\bar{X}^I F_I (X,{\hat A}) - 
\bar{F}_I ({\bar X}, {\bar {\hat A}}) X^I ]$.  This quantity will
play a role in the following.

The associated (Wilsonian) Lagrangian
describing the coupling of these vector multiplets
to supergravity 
is quite complicated \cite{deWit1}.  
Here we only display those terms
which will be relevant for the computation of the entropy of
a static supersymmetric black hole,
\be
{8 \pi \cal L} = - \ft{1}{2} e^{-{\cal K}}R 
+ \ft12( {i} F_{\hat{A}} \,\hat{C} +
\mbox{h.c.}) + \cdots \,,
\label{Paction}
\eq
where ${\hat C} = 64 \,C^{- \mu \nu \rho \sigma } 
C^-_{ \mu \nu \rho \sigma}
+ 16 \,\varepsilon_{ij}\, T^{\mu \nu i j} f_{\mu}{}^{\rho\,} T_{\rho \nu k l}
\,\varepsilon^{kl} + \cdots \,$.  Here $C^-_{ \mu \nu \rho \sigma}$
denotes the anti-selfdual 
part of the Weyl tensor $C_{ \mu \nu \rho \sigma}$, and
$f_{\mu}{}^{\nu} = \ft{1}{2} R_{\mu}{}^{\nu} - \ft{1}{12} R \,
\delta_{\mu}{}^{\nu}+ \cdots$.  Eventually we will set $e^{-{\cal K}}$
to unity 
in order to obtain a properly normalized Einstein-Hilbert term.  In this
way we fix the local scale invariance which is present in a superconformal
formulation of the theory.
We note that the Lagrangian contains $C_{ \mu \nu \rho
\sigma}^2$-terms, but no terms 
involving derivatives of the Riemann curvature tensor.  By expanding the
holomorphic function $F(X, {\hat A})$ in powers of ${\hat A}$, 
$F(X, {\hat A}) =\sum_{g=0}^{\infty} F^{(g)}(X) {\hat A}^g$, we see that
the Lagrangian (\ref{Paction}) contains an infinite set of higher-derivative
curvature terms of the type $C^2{}_{\!\!\!\!\mu \nu \rho \sigma} 
( T^{ab\, i j} \varepsilon_{ij}){}^{2g-2}$ ($g\geq 1$) with
field-dependent coupling functions $F^{(g)} (X)$.

As alluded to above, we will view a static supersymmetric black hole
solution of the Lagrangian (\ref{Paction}) as a solitonic interpolation
between two $N=2$ supersymmetric groundstates, namely flat spacetime at
spatial infinity and the horizon, whose geometry
we now proceed to determine.  The spacetime line element associated with
a static spherically symmetric solution is of the form 
(\ref{StaticMetric}) in isotropic
coordinates.  The near-horizon solution can be specified by imposing 
full $N=2$ supersymmetry on the solution.  This is achieved
by requiring that the supersymmetry variations of all the fermions present
in the theory vanish in the bosonic black hole background.  
We stress here that we do not analyze the
equations of motion, as their validity is implied by full $N=2$
supersymmetry. 
A careful
analysis \cite{CarDeWMoh} 
of the resulting restrictions on the bosonic background then shows 
that the $X^I$ and ${\hat A}$ are constant at the horizon.
Since the
black hole is charged and can carry both 
magnetic and electric charges $(p^I, q_I)$,
the associated quantity $Z$ (\ref{z}) is therefore generically 
non-vanishing and constant at the horizon.
Moreover, 
we also find that 
$T^{01\,ij} = - i\, T^{23\,ij} = 2
\varepsilon^{ij} \, e^{-{\cal K}/2}\, \bar Z^{-1}$,
while all other
components of $T^{ab\,ij}$ 
vanish. Therefore we have $\hat A =-  64\,e^{-{\cal K}}\,\bar Z^{-2}$. 
And finally,
the near-horizon spacetime geometry is determined
to  be of the Bertotti-Robinson type,
that is of the form (\ref{StaticMetric}) with $e^{2g(r)} = e^{-2f(r)}=
e^{-{\cal K}} \, \vert Z\vert^{-2} \,r^2$.

The requirement of $N=2$ supersymmetry at the horizon does not by itself
fix the actual values of the constants $X^I$.  To do so,
we have to invoke the so-called fixed-point behaviour \cite{FerKalStr}
for the scalar fields $X^I$
at the horizon (for a recent reference on the fixed-point behaviour, see
\cite{Moore}).  The fixed-point behaviour implies that regardless of 
what the values of the scalar fields are at spatial infinity, they 
always take the same
values at the horizon.  In the
absence of higher-derivative terms 
it has been shown \cite{FerKalStr} that the supersymmetric black hole
solutions do indeed always exhibit
fixed-point behaviour as a result of their residual $N=1$ supersymmetry.
In the presence of higher-derivative terms this has
not yet been shown to be the case.  In the following we will assume 
that such a fixed-point
behaviour holds for black hole solutions of the Lagrangian (\ref{Paction}).
{From} electric-magnetic duality one may then deduce that the values of
the scalar fields $X^I$ at the horizon are  determined from a set
of equations, called the stabilization equations, which take the
following form \cite{BCDWLMS,CarDeWMoh}
\be
\bar{Z} \left( \begin{array}{c} X^I \\ F_I (X,{\hat A}) \\ \end{array} \right)
- Z \left( \begin{array}{c} \bar{X}^I \\ \bar{F}_I ({\bar X}, {\bar {\hat A}})
 \\ \end{array} \right)
= i 
e^{-{\cal K}/2}
\left( \begin{array}{c} p^I \\ q_I \\ \end{array} \right) \,.
\label{stab}
\eq
At this point it is convenient to introduce rescaled variables $Y^I = 
e^{{\cal K}/2}
\bar{Z} X^I$ and $\Upsilon = e^{ {\cal K} } 
\bar{Z}^2 \hat{A} = -64$. Using the homogeneity property of $F$
mentioned earlier, it follows that 
the stabilization equations (\ref{stab})
now simply read
$Y^I - \bar{Y}^I = ip^I$ and $F_I(Y,\Upsilon) - \bar{F}_I(\bar{Y}, 
\bar{\Upsilon}) = i q_I$.  These equations then determine the value of
the rescaled fields $Y^I$ in terms of the charges carried by the
black hole.  On the other hand, it follows 
from (\ref{z}) that $|Z|^2 =
p^I F_I(Y,\Upsilon) - q_I Y^I$, which determines the value of $|Z|$
in terms of $(p^I, q_I)$. 

The entropy of the static black hole solution described above
can now be computed from (\ref{graventro}), using (\ref{Paction}).  
The result takes the remarkably concise form \cite{CarDeWMoh}
\be
{\cal S} = \pi \left[ |Z|^2 - 256\, \mbox{Im} \,F_{\hat A} \right] \,.
\label{entropia}
\eq
The first term denotes the Bekenstein-Hawking entropy contribution, whereas
the second term is due to Wald's modification of the definition of
the entropy in the presence of higher-derivative terms. Here we
point out that this contribution does not actually originate from the 
$C^2{}_{\!\!\!\!\mu \nu \rho \sigma}$-terms
in the action, because the Weyl tensor vanishes at the horizon, but
from the term in ${\hat C}$ (see below (\ref{Paction})) proportional to the 
product of the Ricci tensor with the tensor field $T^{ab\,ij} T_{cd \,kl}$!
Note that when switching on higher-derivative interactions the value of
$|Z|$ changes and hence 
also the horizon area changes.  There are thus 
two ways in which the presence of 
higher-derivative interactions
modifies the black hole entropy, 
namely by a change of the near-horizon geometry and by an explicit
deviation from the Bekenstein-Hawking area law.
Also note that the entropy (\ref{entropia}) is entirely determined
in terms of the charges carried by the black hole, ${\cal S} = {\cal
S} (q,p)$. 

Let us now exhibit the generic dependence of the macroscopic entropy
(\ref{entropia}) on the charges.  
Let $Q$ denote a generic electric or magnetic charge carried by the 
black hole.  For large charges $Q$, the stabilization equations 
(\ref{stab}) and
the homogeneity property of $F$ imply that the generic
dependence of the entropy (\ref{entropia}) on the charges
is given by 
\beqa
{\cal S} = \pi \; \sum_{g=0}^{\infty} a_g \; Q^{2-2g} \;\;\;
\label{sf}
\eeqa
with constant coefficients $a_g$.
We recall that (\ref{sf}) encodes the contributions from a particular
set of $R^2$-terms, namely terms proportional to 
$C_{\mu \nu \rho \sigma}^2$.  In general, however, the
effective Lagrangian will not only contain these particular terms,
but also many other
(even higher-derivative) curvature terms which, in principle, will lead to
further contributions to the entropy.  One might then worry that the
inclusion of such contributions could wash out the contributions 
appearing in (\ref{sf}).  In the context of string theory this
appears to be unlikely, given that the
coefficients multiplying the various powers of the charges in 
(\ref{sf}) encode topological information about the $N=2$ compactification.

\section{Examples}
Let us now briefly discuss various classes of black hole solutions arising
in string theory compactifications.
We will use the rescaled variables $Y^I = e^{{\cal K}/2}{\bar Z} X^I,
\Upsilon = e^{{\cal K}}{\bar Z}^2 {\hat A} = - 64$ throughout.

Let us first consider 
type-IIA string theory compactified on a Calabi-Yau
threefold, in the limit where the volume of the Calabi-Yau threefold
is taken to be large.  For the associated homogenous function
$F(Y,\Upsilon)$ we take (with $I = 0, \ldots, n$ and  $A = 1,
\ldots, n$) 
\be
F(Y,\Upsilon) 
= \frac{D_{ABC} Y^A Y^B Y^C}{Y^0} + 
d_{A}\, \frac{Y^A}{Y^0} \; \Upsilon
\,\,\;,\,\,\; D_{ABC} = - \ft16 \,  C_{ABC} \,\,\;,\,\,\;
d_{A} = - \ft{1}{24} \, \ft{1}{64}\; c_{2A} \,,
\label{prep2a}
\eq
where the coefficients $C_{ABC}$ denote the intersection numbers
of the four-cycles 
of the Calabi-Yau threefold,  
whereas the coefficients ${c_{2A}}$ denote its second Chern-class
numbers \cite{Bershadskyetal}.  The Lagrangian (\ref{Paction})
associated with 
this homogenous function thus contains a term proportional to 
$c_{2A} \, {\rm Im }\, z^A \,C_{\mu \nu \rho \sigma}^2$, where 
$z^A = {Y^A}/{Y^0}$.
The associated stabilization
equations can be solved \cite{CarDeWMoh}
for black holes with $p^0 = 0$,
yielding $Y^I=Y^I(q,p)$. The result for the 
macroscopic entropy, which is computed from 
(\ref{entropia}), reads \cite{CarDeWMoh}
\beqa
{\cal S}= 
2 \pi \sqrt{\ft16 \, |{\hat q}_0|  (C_{ABC} \,p^A p^B p^C + 
{c}_{2A} \, p^A) }\,,
\label{entrot2}
\eeqa
where ${\hat q}_0 = q_0 + \ft{1}{12} D^{AB} q_A q_B \,,\, D_{AB} = D_{ABC} p^C 
\,,\, 
D_{AB} D^{BC} = \delta_A^C$.  The expression (\ref{entrot2})
for the macroscopic entropy is 
in exact agreement with 
the microscopic entropy formula computed in \cite{MalStrWit,Vaf}
via state counting.

Next, let us consider black hole solutions arising 
in heterotic string 
compactifications on $K_3 \times T_2$.  The associated tree-level function 
is given by
\beqa
F(Y,\Upsilon) = - \frac{Y^1 Y^a \eta_{ab} 
Y^b}{Y^0} + c \; \frac{Y^1}{Y^0} \; \Upsilon \;\;\;,\;\; a = 2, \dots, n 
\;\;\;,
\label{hetprep}
\eeqa
where  
the real constants $\eta_{ab}$ and 
$c$ are related to the intersection numbers of two-cycles 
and to the 
second Chern-class number of $K_3$, respectively.
For a function $F(Y, \Upsilon)$ of the form (\ref{hetprep}), the stabilization
equations can be solved in full generality and the resulting expression
for the entropy reads \cite{CDWM}
\beqa
{\cal S} = \pi \; \sqrt{ \langle M,M \rangle \langle N,N \rangle
- ( M \cdot N )^2}
\;  \sqrt{1 - \frac{ 512 \, c \,}{\langle N,N \rangle}}
\;,
\label{hetentro}
\eeqa
where the $M_I = (q_0, - p^1, q_2, q_3, \dots, q_n)$ 
and the $N^I = (p^0, q_1, p^2,
p^3, \dots, p^n)$ now denote
the electric and magnetic charges
carried by the heterotic black hole.  The 
bilinears $\langle M,M \rangle$ and 
$\langle N,N \rangle$ are given by 
\beqa
\langle M,M \rangle = 2 \left( M_0 M_1 + \ft{1}{4} M_a \eta^{ab} M_b \right)
\;\;\;,\;\;\;
\langle N,N \rangle = 2 \left( N^0 N^1 +  N^a \eta_{ab} N^b \right)
\;,
\eeqa
whereas $M \cdot N = M_I N^I$.
These bilinears are invariant under
tree-level target-space duality transformations of the
charges \cite{clm}.
We thus see that turning on a term proportional
to ${\rm Im }\, z^1 \, C_{\mu \nu \rho \sigma}^2$ 
($z^1 = {Y^1}/{Y^0}$) in the Lagrangian
leads again to a modification of the entropy.

There will be various 
corrections to (\ref{entrot2}) and to (\ref{hetentro})
from terms proportional to $C_{\mu \nu \rho \sigma}^2 \,
{\hat A}^{g-1} (g \geq 2)$ in the
effective Lagrangian (\ref{Paction}).  
These corrections are such that they no longer
preserve the square-root feature of the entropy formulae
(\ref{entrot2}) 
and (\ref{hetentro}) \cite{CDWM}.

The examples of black hole solutions discussed so far occur
in $N=2$ compactifications of string theory.  Let us now consider
black hole solutions occuring in 
heterotic $N=4$ compactifications.  At tree-level
there is a term proportional 
to ${\rm Re} \,S \,C_{\mu \nu \rho \sigma}^2$ 
in the effective Lagrangian of heterotic 
string theory compactified on a six-torus, where $S$ here 
denotes the heterotic dilaton field.  The entropy of a black hole
solution in the presence of such an $C_{\mu \nu \rho \sigma}^2$-term
is given by an expression analogous to
(\ref{hetentro}).  As can be seen from (\ref{hetentro}),
the analogous expression is not invariant under the exchange
of the electric and the magnetic charges.  The heterotic $N=4$
theory is, however, expected to be invariant under strong-weak
coupling duality.  Here we recall that it is essential to distinguish
between the Wilsonian couplings of an effective theory and the
physical (in general momentum-dependent) effective couplings.  
In theories with massless fields,
the effective couplings are different
from the Wilsonian couplings and do not share their analytic
properties.  
In the context of heterotic $N=4$ compactifications, the effective
coupling function 
multiplying $C_{\mu \nu \rho \sigma}^2$ is non-holomorphic and
invariant under 
strong-weak coupling duality transformations \cite{HarMoo}, whereas
the Wilsonian coupling function, which at tree-level is
proportional to $S$, 
is holomorphic although not invariant under strong-weak coupling duality.
Thus, 
in order to arrive at a manifestly
strong-weak coupling duality invariant expression for the entropy
of a heterotic $N=4$ black hole, its computation should be based
on the effective rather than the Wilsonian coupling function of 
$C^2_{\mu \nu \rho \sigma}$.

We will now restrict ourselves to black hole solutions in an
$N=2$ subsector of the heterotic $N=4$ theory.  Both the stabilization
equations (\ref{stab}) and the entropy formula
(\ref{entropia}) were derived in the Wilsonian context and as such
they are defined in terms of a holomorphic function $F$.  In the
effective coupling approach, on the other hand, we expect that both
(\ref{stab}) and (\ref{entropia}) 
will receive non-holomorphic corrections so as to be consistent
with strong-weak coupling duality.  The function $F$, in particular,
will not any longer be holomorphic:
\beqa
F(Y, \bar Y, \Upsilon) =  -
 \frac{Y^1 Y^a \eta_{ab} 
Y^b}{Y^0} + \; 
F^{(1)}( z^1, {\bar z}^1 ) 
\;  \Upsilon \;\;\;,
\label{nonholof}
\eeqa
where we require that (\ref{nonholof}) turns into the tree-level
function (\ref{hetprep}) at weak coupling, that is 
$F^{(1)}( z^1, {\bar z}^1 ) \rightarrow
c \,z^1$ as $ S + {\bar S} \rightarrow \infty$, where
$S = -i z^1 = -i \, Y^1/Y^0$.  The associated stabilization
equations now read
\beqa
Y^I - {\bar Y}^I = i p^I \;\;\;,\;\;\; F_I(Y, \bar Y,  \Upsilon) - 
{\bar F}_I (Y, \bar Y, \bar \Upsilon) = i q_I \;\;\;,\;\;\;
\Upsilon = -64 \;\;\;.
\label{nonhstab}
\eeqa
The form of $F^{(1)}( z^1, {\bar z}^1 )$ can then be determined by 
requiring (\ref{nonhstab})
to transform in a consistent way under strong-weak coupling
duality transformations and also
by requiring $F^{(1)}( z^1, {\bar z}^1 )$ to have the weak-coupling
behaviour specified above \cite{CDWM}:
\beqa
 F^{(1)} (S, {\bar S}) = 
- 6 \,i \, \frac{c }{\pi}  \,  \Big(  \log \eta^2 (S)
+ \log (S + {\bar S}) \Big) \,,
\eeqa
where $\eta (S)$ denotes the Dedekind function.
The associated non-holomorphic quantity $|Z|^2 = 
 p^I F_I(Y, {\bar Y}, \Upsilon) - q_I Y^I $ is then
invariant under strong-weak coupling duality.  In analogy to (\ref{entropia})
we thus propose the following
strong-weak coupling duality invariant
expression for the entropy of a black hole in an $N=2$ subsector
of the heterotic $N=4$ theory \cite{CDWM}:
\beqa
{\cal  S} = \pi \Big[ \; |Z|^2 
 - 256\, {\rm Im} \,\Big( \, F^{(1)} (S, {\bar S}) + 3 \, i \,
 \frac{c }{\pi} \, 
 \log (S + {\bar S}) 
 \Big)\; \Big]
\,. 
\label{entropiaynonholo}
\eeqa
The additional non-holomorphic piece in (\ref{entropiaynonholo})
is there to render the expression invariant under strong-weak
coupling duality.
We note that the value of the dilaton $S$ at the horizon 
is, in principle, determined in terms of the charges $M_I$ and $N^I$ carried
by the black hole through the stabilization equations (\ref{nonhstab}), 
although
in practice they are hard to solve.
We refer to \cite{CDWM} for a more detailed discussion
of these and related issues.

\vskip0.5cm

\leftline{\bf Acknowledgements}

\noindent
We thank Soo-Jong Rey for valuable discussions. 
T.M. thanks the Institute for Theoretical Physics of Utrecht University 
for hospitality during the final stages of this work.



\begin{thebibliography}{77}
%
\bibitem{Haw}
S. W. Hawking, {\it Phys. Rev. Lett.} {\bf 26} (1971) 1344;
J. M. Bardeen, B. Carter and S. W. Hawking, 
{\it Commun. Math. Phys.} {\bf 31} (1973) 161.
%
\bibitem{Bek}
J. D. Bekenstein, {\it Phys. Rev.} {\bf D7} (1973) 2333; 
{\it Phys. Rev.} {\bf D9} (1974)
3292;
S. W. Hawking, {\it Comm. Math. Phys.} {\bf 43} (1975) 199.
%
\bibitem{StromVafa}
A. Strominger and C. Vafa, {\it Phys. Lett.} {\bf B379} (1996) 
99, {\tt hep-th/9601029}.

\bibitem{Wald} R. M. Wald, {\it Phys. Rev.} {\bf D48} (1993) 3427, 
{\tt gr-qc/9307038}.



\bibitem{JacKaMy}
T. Jacobson, G. Kang and R. C. Myers, {\it 
Phys. Rev.} {\bf D49} (1994) 6587, {\tt gr-qc/9312023}; 
{\it Black Hole Entropy
in Higher Curvature Gravity}, {\tt gr-qc/9502009}.
%

\bibitem{IWald}
V. Iyer and R. M. Wald, {\it Phys. Rev.} {\bf D50} (1994) 846, 
{\tt gr-qc/9403028}.

\bibitem{WaldRev}
R. M. Wald, {\it Black Holes and Thermodynamics}, 
{\tt gr-qc/9702022}; {\it Gravitation, Thermodynamics, and Quantum Theory},
{\tt gr-qc/9901033}.
\bibitem{Pol}
J. Polchinski, {\it String Theory, Vol. II}, Cambridge University Press (1998).
\bibitem{Mal}
J. M. Maldacena, {\it Black Holes in String Theory}, {\tt hep-th/9607235}.
\bibitem{MalStrWit}
J. M. Maldacena, A. Strominger and E. Witten, 
{\it J. High Energy Phys.} {\bf 12} (1997) 2, 
{\tt hep-th/9711053}.
%
\bibitem{Vaf}
C. Vafa, {\it Adv. Theor. Math. Phys.} {\bf 2} (1998) 207, 
{\tt hep-th/9711067}.
%
\bibitem{CarDeWMoh}
G. L. Cardoso, B. de Wit and T. Mohaupt, 
{\it Corrections to macroscopic supersymmetric black-hole entropy},
to appear in {\it Phys. Lett.} {\bf B}, 
{\tt hep-th/9812082}.
%
\bibitem{HawEll}
S. W. Hawking and G. F. R. Ellis, {\it The Large Scale Structure
of Spacetime}, Cambridge University Press (1973).
%
\bibitem{Carter} B. Carter, in {\it Black Holes}, eds. C. DeWitt
and B. S. DeWitt, Gordon and Breach, New York (1973).
%

\bibitem{Visser} M. Visser, {\it Phys. Rev.} {\bf D48} (1993) 5697, 
{\tt hep-th/9307194}.


\bibitem{Waldbook}
R. M. Wald, {\it General Relativity}, Chicago University Press (1984).




\bibitem{GibHul}
G. W. Gibbons and C. M. Hull, {\it Phys. Lett.} {\bf 109B} (1982) 190;
G. W. Gibbons, in
{\it Supersymmetry, Supergravity and Related Topics}, eds.
F. del Aguila, J. de Azc\'arraga and L. Ib\'an\~ez, World Scientific 
(1985) 147.



\bibitem{RacWal}
I. R\'acz and R. M. Wald, {\it Class. Quant. Grav.} {\bf 9} (1992) 2643; 
{\it 
Global Extensions of Spacetimes Describing Asymptotic Final States of Black
Holes}, {\tt gr-qc/9507055}.

\bibitem{GhoMit}
A. Ghosh and P. Mitra, {\it Phys. Rev. Lett.} {\bf 78} (1997) 1858,
{\tt hep-th/9609006};
P. Mitra, {\it Phys. Lett.} {\bf B441} (1998) 89,
{\tt hep-th/9807094}.
%
\bibitem{KieLou}
C. Kiefer and J. Louko, {\it Annalen Phys.} {\bf 8} (1999) 67, 
{\tt gr-qc/9809005}.
%

\bibitem{IyeWal2}
V. Iyer and R. M. Wald, 
{\it Phys. Rev.} {\bf D52} (1995) 4430,
{\tt gr-qc/9503052}.
%
\bibitem{MajPap}
S. D. Majumdar, {\it Phys. Rev.} {\bf 72} (1947) 930; 
A. Papapetrou, {\it Proc. R. Irish Acad.} {\bf A51} (1947) 191.
%
\bibitem{FerKalStr}
S. Ferrara, R. Kallosh and A. Strominger, {\it Phys. Rev.} {\bf D52} 
(1995) 5412, {\tt hep-th/9508072};
A. Strominger, {\it Phys. Lett.} {\bf B383} (1996) 39, {\tt hep-th/9602111};
S. Ferrara and R. Kallosh,
{\it Phys. Rev.} {\bf D54} (1996) 1514, {\tt hep-th/9602136};
{\it Phys. Rev.} {\bf D54} (1996) 1525, {\tt hep-th/9603090}.
%
\bibitem{BCDWLMS}
K. Behrndt, G. L. Cardoso, B. de Wit, D. L\"ust, T. Mohaupt 
and W. A. Sabra,
{\it Phys. Lett.} {\bf B429} (1998) , {\tt hep-th/9801081}.
%
\bibitem{deWvPr}
B. de Wit, J. W. van Holten and A. Van Proeyen, {\it Nucl. Phys.} {\bf
B167} (1980) 186, 
{\it Nucl. Phys.} {\bf B184} (1981) 77;
B. de Wit and A. Van Proeyen, {\it Nucl. Phys.}
{\bf B245} (1984) 89;  
B. de Wit, P. G. Lauwers and A. Van Proeyen,
{\it Nucl. Phys.} {\bf B255} (1985) 569.
%
\bibitem{BDRDW}
E. Bergshoeff, M. de Roo and B. de Wit, {\it Nucl. Phys.} {\bf B182}
(1981) 173. 
%
%
\bibitem{deWit1}
B. de Wit, {\em Nucl. Phys. (Proc. Suppl.)} {\bf B49} (1996) 191, 
{\tt hep-th/9602060}.
%
\bibitem{Moore} G. Moore, {\it Attractors and Arithmetic}, 
{\tt hep-th/9807056}; {\it Arithmetic and Attractors}, 
{\tt hep-th/9807087}. 
%
\bibitem{Bershadskyetal}
M. Bershadsky, S. Cecotti, H. Ooguri and C. Vafa, 
{\it Nucl. Phys.} {\bf B405} (1993) 279,
{\tt hep-th/9302103};
{\em Comm. Math. Phys.} {\bf  165} (1994) 311, {\tt hep-th/9309140}.
%
\bibitem{CDWM} G. L. Cardoso, B. de Wit and T. Mohaupt, to appear.
\bibitem{clm} G. L. Cardoso, D. L\"ust and T. Mohaupt, 
{\it Phys. Lett.} {\bf B388} (1996) 266, {\tt hep-th/9608099}.
\bibitem{HarMoo} J. A. Harvey  and G. Moore, 
{\it Phys. Rev.} {\bf D57} (1998) 2323, {\tt hep-th/9610237}.


%

%
%



\end{thebibliography}
\end{document}